\documentclass[preprint,12pt]{elsarticle}

\usepackage{amssymb}
\usepackage{amsmath,graphicx,amsfonts,fancybox,color,amssymb,lmodern,soul}
\usepackage[T1]{fontenc}
\usepackage{pgfplots}
\usepackage{algpseudocode}
\usepackage{algorithm}
\pgfplotsset{compat=1.5}

\newcommand{\mbf}{\mathbf}

\newcommand{\ssi}[1]{\textcolor{red}{#1}}
\newcommand{\hspi}[1]{\textcolor{red}{#1}}

\usepackage{bm}

\def\Cbold{\mathbf{C}}

\def\Ebold{\mathbf{E}}

\def\fbold{\mathbf{f\/}} 
 
\def\gbold{\mathbf{g}}

\def\Kbold{\mathbf{K}} 
\def\kbold{\mathbf{k}}

\def\Mbold{\mathbf{M}} 
\def\mbold{\mathbf{m}}

\def\ubold{\mathbf{u}}

\def\zerobold{\bm{0}}

\def\Phibold{\bm{\Phi}}

\def\pibold{\bm{\pibold}}

\def\boxed#1{\setbox0=\hbox{$\displaystyle{#1}$}\hbox{\lower.6pt\hbox{\lower
 3pt\hbox{\lower 1\dp0\hbox{\vbox{\hrule height .6pt \hbox{\vrule width
 .6pt \hskip 3pt\vbox{\vskip 3pt\box0\vskip3pt}\hskip 3pt \vrule width
    .6pt}\hrule height .6pt}}}}}}
\def\dddot#1{{\mathop{#1}\limits^{\vbox to -1.4pt{\kern-2pt
   \hbox{\tenrm...}\vss}}}}
\def\ddddot#1{{\mathop{#1}\limits^{\vbox to -1.4pt{\kern-2pt
   \hbox{\tenrm....}\vss}}}}

\def\half{{\textstyle{1\over2}}}

\def\squeezedmatrixrt#1{\null\,\vcenter{\normalbaselines\m@th
    \ialign{\hfil$##$&&$\quad\!\!\!$\hfil$##$\crcr
      \mathstrut\crcr\noalign{\kern-\baselineskip}
      #1\crcr\mathstrut\crcr\noalign{\kern-\baselineskip}}}\,}
\def\crampedmatrix#1{\null\,\vcenter{\normalbaselines\m@th
    \ialign{\hfil$##$\hfil&&$\,\,$\hfil$##$\hfil\crcr
      \mathstrut\crcr\noalign{\kern-\baselineskip}
      #1\crcr\mathstrut\crcr\noalign{\kern-\baselineskip}}}\,}
\def\crampedmatrixrt#1{\null\,\vcenter{\normalbaselines\m@th
    \ialign{\hfil$##$&&$\,\,$\hfil$##$\crcr
      \mathstrut\crcr\noalign{\kern-\baselineskip}
      #1\crcr\mathstrut\crcr\noalign{\kern-\baselineskip}}}\,}

\def\bordertable#1{\begingroup \m@th
  \setbox\z@\vbox{\def\cr{\crcr\noalign{\kern2\p@\global\let\cr\endline}}%
    \ialign{$##$\hfil\kern2\p@\kern\p@renwd&\thinspace\hfil$##$\hfil
      &&\quad\hfil$##$\hfil\crcr
      \omit\strut\hfil\crcr\noalign{\kern-\baselineskip}%
      #1\crcr\omit\strut\cr}}%
  \setbox\tw@\vbox{\unvcopy\z@\global\setbox\@ne\lastbox}%
  \setbox\tw@\hbox{\unhbox\@ne\unskip\global\setbox\@ne\lastbox}%
  \setbox\tw@\hbox{$\kern\wd\@ne\kern-\p@renwd\left.\kern-\wd\@ne
    \global\setbox\@ne\vbox{\box\@ne\kern2\p@}%
    \vcenter{\kern-\ht\@ne\unvbox\z@\kern-\baselineskip}\,\right.$}%
  \null\;\vbox{\kern\ht\@ne\box\tw@}\endgroup}

\def\squeezedmorematrix#1{\null\,\vcenter{\normalbaselines\m@th
    \ialign{\hfil$##$\hfil&&$\quad\!\!\!\!$\hfil$##$\hfil\crcr
      \mathstrut\crcr\noalign{\kern-\baselineskip}
      #1\crcr\mathstrut\crcr\noalign{\kern-\baselineskip}}}\,}

\def\squeezedlessmatrix#1{\null\,\vcenter{\normalbaselines\m@th
    \ialign{\hfil$##$\hfil&&$\quad\!$\hfil$##$\hfil\crcr
      \mathstrut\crcr\noalign{\kern-\baselineskip}
      #1\crcr\mathstrut\crcr\noalign{\kern-\baselineskip}}}\,}

\def\btop#1{{\mathop{#1}\limits^{\vbox to -1.4pt{\kern-3pt
   \hbox{$\scriptscriptstyle\bullet$}\vss}}}}
\def\bbtop#1{{\mathop{#1}\limits^{\vbox to -1.4pt{\kern-3pt
   \hbox{$\scriptscriptstyle\bullet\bullet$}\vss}}}}
\catcode`@=12 

\chardef\@=`\@
\def\wiggle{\lower3pt\hbox{\char`\~}}

\chardef\caret=`\^
\def\sqr#1#2{{\vcenter{\hrule height.#2pt
           \hbox{\vrule width.#2pt height#1pt \kern#1pt
           \vrule width.#2pt}
           \hrule height.#2pt}}}

\mathchardef\triangledown="0235

\mathchardef\varGamma="0100
\mathchardef\varDelta="0101
\mathchardef\varTheta="0102
\mathchardef\varLambda="0103
\mathchardef\varXi="0104
\mathchardef\varPi="0105
\mathchardef\varSigma="0106
\mathchardef\varUpsilon="0107
\mathchardef\varPhi="0108
\mathchardef\varPsi="0109
\mathchardef\varOmega="010A
\mathchardef\varkappa="017E

\def\overset#1\to#2{{\mathop{#2}^{#1}}}
\def\underset#1\to#2{{\mathop{#2}_{#1}}}
\def\oversetbrace#1\to#2{{\overbrace{#2}^{#1}}}
\def\undersetbrace#1\to#2{{\underbrace{#2}_{#1}}}

\def\@{\char'100 }

\journal{CMAME}

\begin{document}

\begin{frontmatter}


\title{A Staggered Explicit-Implicit Finite Element Formulation for Electroactive Polymers}

\author[ss]{Saman Seifi}
\author[kcp]{K.C. Park}
\author[hsp]{Harold S. Park\corref{cor1}}
\ead{parkhs@bu.edu}

\cortext[cor1]{Corresponding Author}

\address[ss]{Department of Mechanical Engineering, Boston University, Boston, MA 02215}
\address[kcp]{Department of Aerospace Engineering, University of Colorado at Boulder, Boulder, CO 80309}
\address[hsp]{Department of Mechanical Engineering, Boston University, Boston, MA 02215}

\begin{abstract}

Electroactive polymers such as dielectric elastomers (DEs) have attracted significant attention in recent years.  Computational techniques to solve the coupled electromechanical system of equations for this class of materials have universally centered around fully coupled monolithic formulations, which while generating good accuracy requires significant computational expense. However, this has significantly hindered the ability to solve large scale, fully three-dimensional problems involving complex deformations and electromechanical instabilities of DEs.  In this work, we provide theoretical basis for the effectiveness and accuracy of staggered explicit-implicit finite element formulations for this class of electromechanically coupled materials, and elicit the simplicity of the resulting staggered formulation.  We demonstrate the stability and accuracy of the staggered approach by solving complex electromechanically coupled problems involving electroactive polymers, where we focus on problems involving electromechanical instabilities such as creasing, wrinkling, and bursting drops.  In all examples, essentially identical results to the fully monolithic solution are obtained, showing the accuracy of the staggered approach at a  significantly reduced computational cost.

\end{abstract}

\begin{keyword}
staggered \sep explicit-implicit \sep creasing \sep dielectric elastomer \sep wrinkling \sep surface tension

\end{keyword}

\end{frontmatter}

% \linenumbers

\section{Introduction} 

Dielectric elastomers (DEs) have attracted significant attention in recent years as a soft and flexible actuation material~\citep{carpiSCIENCE2010,brochuMRC2010,biddissMEP2008,zhaoAPR2014}.  They have been found to provide excellent overall performance in actuation-based applications, including high specific elastic energy density, good efficiency and high speed of response.  Furthermore, DEs are typically lightweight, flexible and inexpensive materials which makes them ideal candidates for high performance, low cost applications where fabrication of the DEs into a wide range of shapes and structures can easily be realized~\cite{zhangAEM2005}.  While DEs have been found to exhibit good performance with respect to a variety of actuation-relevant properties, including strain, actuation pressure, efficiency, response speed, and density~\cite{pelrineSAA1998}, the key source of the technological excitement surrounding DEs stems from the fact that if sandwiched between two compliant electrodes that apply voltage to the elastomer, the DE can exhibit both significant thinning and in-plane expansion, where the in-plane expansion can often exceed several hundred percent~\citep{keplingerSM2012}.  The ability to undergo such large deformations has led to DEs being studied for both actuation-based applications, including artificial muscles and flexible electronics, and also for generation-based applications and energy harvesting~\citep{carpiSCIENCE2010,brochuMRC2010,mirMT2007}. 

Various computational formulations for DEs based on the finite element method (FEM) have emerged in the past decade~\citep{parkIJSS2012,parkSM2013,parkCMAME2013,zhouIJSS2008,zhaoAPL2007,vuIJNME2007,wisslerSMS2005,buschelIJNME2013,khanCM2013,henannJMPS2013,liSMS2012,wangJMPS2016,schloglCMAME2016,seifiIJSS2016,yvonnetCMAME2017}.  While these differ depending on various factors, including the field theory they are formulated on, whether they account for material effects such as viscoelasticity, or whether they are quasi-static or dynamic, nearly all of them have been solved using a fully coupled, monolithic formulation.  While the monolithic formulation ensures the correct electromechanical coupling, it comes with significant computational expense, and as such nearly all computational examples involving DEs have been on two-dimensional (2D) problems because of the additional degree of freedom the electrostatic problem adds to the structural problem for each spatial dimension.  

We mention a related work by Zhang and co-workers~\cite{zhangIJCM2014} that simply applied an explicit-implicit   computational procedure to an analysis of DEs.  However, no reference or discussion on the stability and accuracy aspects of partitioned explicit-implicit procedures~\cite{belytschkoIJNME1978,hughesJAM1978,parkJAM1980} was provided; hence, they offered no rationale for its stability restrictions and accuracy analysis.  Moreover, they did not report any comparison  of their work to a series of reported benchmark monolithic solutions obtained by fully implicit-implicit procedures~\cite{parkSM2013,parkCMAME2013,seifiIJSS2016}.   As a result, it is unclear as to the applicability ranges, effectiveness, and overall potential for staggered methods in addressing electromechanically coupled phenomena in electroactive polymers like DEs.  In passing, we also note that the possibility of uncoupling the electrostatic and structural fields has also been discussed in multiple works~\cite{vuIJNME2007,schloglCMAME2016}.  Nevertheless, a staggered solution methodology in general and a corresponding investigation of its robustness, stability and accuracy, particularly for electromechanical instabilities, remains lacking.  

In the present work, we present a staggered explicit-implicit finite element formulation for DEs, complete with a criterion for selecting stable step sizes and an accuracy assessment. Specifically, the structural problem is solved explicitly, while the electrostatic problem is solved implicitly. From a historical perspective, our algorithm may be akin to a node-by-node partition of Belytschko and Mullen~\cite{belytschkoIJNME1978}, although both the structural system and the dielectric field equations occupy the same spatial domain.  We demonstrate the  robustness of the present algorithm in  solving problems involving complex electromechanical instabilities, including creasing and wrinkling~\cite{parkCMAME2013,seifiIJSS2016,wangPRE2013,wangPRL2011a}, bursting drops in dielectric solids~\cite{parkSM2013,wangNC2012}, and 3D problems.  In all cases, the staggered methodology provides effectively identical results as a previous dynamic, fully coupled monolithic formulation~\cite{parkCMAME2013}, though for a significantly reduced computational cost.  

\section{Fully Coupled, Monolithic Formulation}

\subsection{Field and Constitutive Equations}

The numerical results we present in this work are based upon a FE discretization of the electromechanical field theory proposed by Suo and co-workers~\citep{suoJMPS2008,suoAMSS2010}.  This fully coupled, monolithic FE formulation has been described in previous works~\cite{parkIJSS2012,parkSM2013,parkCMAME2013}, and so we will briefly outline the relevant background here while referring the interested reader to previous works for further details.

In this field theory at mechanical equilibrium, the nominal stress $S_{iJ}$ satisfies the following (weak) equation:
\begin{equation}\label{eq:suo1} \int_{V}S_{iJ}\frac{\partial\xi_{i}}{\partial X_{J}}dV=\int_{V}\left(B_{i}-\rho\frac{\partial^{2}x_{i}}{\partial t^{2}}\right)\xi_{i}dV+\int_{A}T_{i}\xi_{i}dA,
\end{equation}
where $\xi_{i}$ is an arbitrary vector test function, $B_{i}$ is the body force per unit reference volume $V$, $\rho$ is the mass density of the material and $T_{i}$ is the force per unit area that is applied on the surface $A$ in the reference configuration.  

For the electrostatic problem, the nominal electric displacement $\tilde{D}_{I}$ satisfies the following (weak) equation:
\begin{equation}\label{eq:suo2} -\int_{V}\tilde{D}_{I}\frac{\partial\eta}{\partial X_{I}}dV=\int_{V}q\eta dV+\int_{A}\omega\eta dA,
\end{equation}
where $\eta$ is an arbitrary scalar test function, $q$ is the volumetric charge density and $\omega$ is the surface charge density, both with respect to the reference configuration.  It can be seen that the strong form of the mechanical weak form in (\ref{eq:suo1}) is the momentum equation, while the strong form of the electrostatic weak form in (\ref{eq:suo2}) is Gauss's law.  

As the governing field equations in (\ref{eq:suo1}) and (\ref{eq:suo2}) are decoupled, the electromechanical coupling occurs through the material laws.  The hyperelastic material law we adopt here has been utilized in the literature to study the nonlinear deformations of electrostatically actuated polymers; see the works of~\citet{vuIJNME2007}, and~\citet{zhaoAPL2007}.  Due to the fact that the DE is a rubber-like polymer, phenomenological free energy expressions are typically used to model the deformation of the polymer chains.  In the present work, we will utilize the form~\citep{vuIJNME2007,zhaoAPL2007}
\begin{equation}\label{eq:de1} W(\mbf{C},\tilde{\mbf{E}})=\mu W_{0}+\frac{1}{2}\lambda(\ln{J})^{2}-2\mu W_{0}'(3)\ln{J}-\frac{\epsilon}{2}JC_{IJ}^{-1}\tilde{E}_{I}\tilde{E}_{J},
\end{equation}
where $W_{0}$ is the mechanical free energy density in the absence of an electric field, \hspi{$W_{0}'$ is the derivative of $W_{0}$ with respect to the invariant $I_{1}$}, $\epsilon$ is the permittivity, $J=\det(\mbf{F})$, where $\mbf{F}$ is the continuum deformation gradient, $C_{IJ}^{-1}$ are the components of the inverse of the right Cauchy-Green tensor $\mbf{C}$, $\lambda$ is the bulk modulus and $\mu$ is the shear modulus.  The second and third terms in (\ref{eq:de1}) are used to enforce material incompressibility by taking a large ratio of the bulk to the shear modulus $\lambda/\mu$.

We model the mechanical behavior of the DE using the Arruda-Boyce rubber hyperelastic function~\citep{arrudaJMPS1993}, where the mechanical free energy $W_{0}$ in (\ref{eq:de1}) is approximated by the following truncated series expansion,
\begin{eqnarray}\label{eq:de2} W_{0}(I_{1})=\frac{1}{2}(I_{1}-3)+\frac{1}{20N}(I_{1}^{2}-9)+\frac{11}{1050N^{2}}(I_{1}^{3}-27) \\ \nonumber
+\frac{19}{7000N^{3}}(I_{1}^{4}-81)+\frac{519}{673750N^{4}}(I_{1}^{5}-243),
\end{eqnarray}
where $N$ is a measure of the cross link density, $I_{1}$ is the trace of $\mbf{C}$, and where the Arruda-Boyce model reduces to a Neo-Hookean model if $N\rightarrow\infty$.  We note that previous experimental studies of~\citet{wisslerSAA2007a} have validated the Arruda-Boyce model as being accurate for modeling the large deformation of DEs.  

\subsection{Nonlinear, Monolithic Finite Element Model}

The FE model we use was previously developed in~\citep{parkIJSS2012,parkSM2013,parkCMAME2013}.  In that work, the corresponding author and collaborators developed a nonlinear, dynamic FEM formulation of the governing electromechanical field equations of~\citet{suoJMPS2008} in (\ref{eq:suo1}) and (\ref{eq:suo2}).  By using a standard Galerkin FE approximation to both the mechanical displacement and electric potential fields, both static and dynamic FE formulations were obtained.  The static formulation results in the following FE equations~\cite{parkIJSS2012}
\begin{equation}\label{eq:fe1} \left\{\begin{array}{cc}{\Delta\mbf{u}} \\ {\Delta\Phi}\end{array}\right\}=-\left[\begin{array}{cc} {\mbf{K}_{mm}} & {\mbf{K}_{me}} \\ {\mbf{K}_{em}} & {\mbf{K}_{ee}} \end{array}\right]^{-1}\left\{\begin{array}{cc}{\mbf{R}_{m}} \\ {\mbf{R}_{e}}\end{array}\right\}
\end{equation}
If inertial effects in the mechanical momentum are accounted for, an implicit, fully coupled, monolithic nonlinear dynamic FE formulation was obtained with the governing equations~\citep{parkIJSS2012}
\begin{equation}\label{eq:fe2} \left\{\begin{array}{cc}{\Delta\mbf{a}} \\ {\Delta\Phi}\end{array}\right\}=-\left[\begin{array}{cc} {\mbf{M}+\beta\Delta t^{2}\mbf{K}_{mm}} & {\mbf{K}_{me}} \\ {\beta\Delta t^{2}\mbf{K}_{em}} & {\mbf{K}_{ee}} \end{array}\right]^{-1}\left\{\begin{array}{cc}{\mbf{R}_{m}} \\ {\mbf{R}_{e}}\end{array}\right\}
\end{equation}
where $\Delta\mbf{a}$ is the increment in mechanical acceleration, $\Delta\mbf{u}$ is the increment in displacement, $\Delta\Phi$ is the increment in electrostatic potential, $\beta=0.25$ is the standard Newmark time integrator parameter, $\mbf{R}_{m}$ is the mechanical residual, $\mbf{R}_{e}$ is the electrical residual, and the various stiffness matrices $\mbf{K}$ include the purely mechanical ($\mbf{K}_{mm}$), mixed electromechanical ($\mbf{K}_{me}=\mbf{K}_{em}$), and purely electrostatic ($\mbf{K}_{ee}$) contributions.  Details regarding the residual vectors and the various mechanical, electromechanical and electrostatic stiffnesses can be found in previous work~\citep{parkIJSS2012}, and where volumetric locking due to incompressible material behavior was alleviated using the Q1P0 method of~\citet{simoCMAME1985}.  \hspi{We note, as shown in Simo et al.~\cite{simoCMAME1985}, that no additional degrees of freedom or changes in quadrature points are needed as a result of the Q1P0 formulation.} The dynamic formulation was primarily used in previous works~\citep{parkIJSS2012,parkSM2013,parkCMAME2013,seifiIJSS2016,seifiSM2017} due to its ability to capture the evolution and post-instability response for electromechanical instabilities.

\section{Staggered Formulation}

Multiphysics problems may be classified into two categories.  In the first case, each field occupies separate spatial domains as in fluid-structure interaction, where coupling occurs along the spatial boundaries.  In the second case, the coupled interaction fields occupy the same spatial domain, as in electrodynamics and flexible solids, and thermoelastic problems.  Staggered solution methods~\cite{parkASME1977,felippaCMAME2001}  were initially developed for implicit-implicit staggered solutions of fluid-structure interaction problems, then extended to staggered implicit-implicit solutions of thermoelastic problems~\cite{farhatCMAME1991} and electrodynamics interacting with flexible structures~\cite{parkJMS2004}. 

However, those problems are characterized as stiff problems with mild nonlinearities.  For problems undergoing local/global bifurcation and rapidly varying severe nonlinearities,  it is generally agreed that explicit  integration is preferred in order to capture the rapidly varying nonlinearities.  It is for this reason that we will employ the explicit integration method for advancing the solid equations while implicitly solving the electrostatic field equation, viz., an explicit-implicit staggered procedure.

\subsection{Explicit-Implicit Staggered Formulation}

We begin with the momentum equation for the mechanical problem in (\ref{eq:suo1}).  The FE discretization of the momentum equation in (\ref{eq:suo1}) leads to the following nonlinear dynamical equations 
\begin{gather}
\begin{split}
 \Mbold \ddot\ubold       &= \fbold_{ext}  -  \fbold_{int}  \\  
 & \fbold_{ext}    = \int B_i   N_a \ dV  + \int T_i  N_a \ dA \\
   & \fbold_{int}    = \int \bar{S}_{iJ}(\mbf{u}, \tilde{\mbf{E}}) \frac{\partial N_a}{\partial X_J} \ dV        
\label{eq:ssolids}
\end{split}
\end{gather}
where $B_{i}$ is the body force, $T_{i}$ is the traction, and $\bar{S}_{ij}$ is the nominal stress, which is a function of both the mechanical displacements $\mbf{u}^{n}$ and the electric field $\tilde{\mbf{E}}^{n}$ at timestep $n$, and which is obtained as 
\begin{equation}\label{eq:qp5} \bar{\mbf{S}}=\frac{\partial\tilde{W}(\mbf{C},\tilde{\mbf{E}})}{\partial\mbf{C}}|_{\mbf{C}=\Theta^{2/3}\hat{\mbf{C}}}
\end{equation}
where $\hat{\mbf{C}}=J^{-2/3}\mbf{C}$ and with this modification the energy density function is written $\tilde{W}(\Theta^{2/3}\hat{\mbf{C}},\tilde{\mbf{E}})=W$ and in the continuous case $\Theta(\mbf{X},t)=J(\mbf{X},t)$ is a new kinematic variable due to the Q1P0 approach to relieving volumetric locking by Simo \emph{et al.}~\cite{simoCMAME1985}

The FE-discretized mechanical equation in (\ref{eq:ssolids}) can be integrated explicitly in time using the standard central difference time integration algorithm~\cite{hughes1987,blm2002}.  Specifically, $\ddot{\mbf{u}}$ is obtained from Eq. (\ref{eq:ssolids}), at which point the velocity $\dot{\mbf{u}}^{n+\frac{1}{2}}$ and then updated displacement $\mbf{u}^{n+1}$ can be obtained.  This time marching procedure can be written as
\begin{align}\begin{split}
\ddot{\mbf{u}}^{n}&=\mbf{M}_{uu}^{-1}\mbf{f}_{m}^{n} \\
\dot{\mbf{u}}^{n+\frac{1}{2}}&=\dot{\mbf{u}}^{n-\frac{1}{2}}+\Delta t\ddot{\mbf{u}}^{n} \\
\mbf{u}^{n+1}&=\mbf{u}^{n}+\Delta t\dot{\mbf{u}}^{n+\frac{1}{2}}
\label{eq:explicit}
\end{split} \end{align}
where $\mbf{M}_{uu}$ is the mass matrix for the structural problem, $\mbf{f}_{m}^{n}$ is the difference between the external and internal mechanical forces at timestep $n$, and $\ddot{\mbf{u}}^{n}$ is the acceleration at timestep $n$.  Once the FE displacements have been updated to $\mbf{u}^{n+1}$ through the central difference time integration in (\ref{eq:explicit}), the updated voltage $\mbf{\Phi}^{n+1}$ is obtained by solving the following FE discretization of the electrostatic equations in (\ref{eq:suo2}):
\begin{gather} \label{eq:selectric}
\begin{split}
&   \Kbold^{n+1}_{ee}  \ \Phibold^{n+1} = \hat \fbold_e, 
\quad \hat \fbold_{e} = \int q N_a \ dV + \int \omega N_a\ dA 
\end{split}
\end{gather}
where $q$ is the volumetric charge density, $\omega$ is the surface charge density, and where the fully nonlinear deformation-dependent electrostatic stiffness matrix $\mbf{K}^{n+1}_{ee}$ is used to solve the electrostatic equations (\ref{eq:selectric}).  $\mbf{K}^{n+1}_{ee}$ is dependent on the updated displacement $\mbf{u}^{n+1}$ as
\begin{equation}\label{eq:selectric2} \mbf{K}^{n+1}_{ee}=\int \frac{\partial N_{a}}{\partial X_{J}}\epsilon J\mbf{C}^{-1}(\mbf{u}^{n+1})\frac{\partial N_{b}}{\partial X_{L}}dV
\end{equation}
Once the converged voltage $\mbf{\Phi}^{n+1}$ has been obtained at timestep $n+1$, the staggered procedure begins again with the solution of the mechanical momentum equation.  The entire staggered explicit-implicit procedure is detailed in \ssi{Algorithm} \ref{alg:euclid} below.  

\algnewcommand{\algorithmicgoto}{\textbf{go to}}%
\algnewcommand{\Goto}[1]{\algorithmicgoto~\ref{#1}}%
\begin{algorithm}
  \caption{Flowchart of Staggered Explicit-Implicit Formulation }\label{alg:euclid}
  \begin{algorithmic}[1]
    \Procedure{Initialization}{}
    \State Set the initial conditions $\mathbf{u}^{0}$, $\dot{\mathbf{u}}^{0}$,$\Phibold^{0}$ and compute $\mathbf{M}$
    \State Compute $\mathbf{F}^0=\mathbf{I}+\nabla_{\mathbf{X}}\mathbf{u}^0$ and $\tilde{\mathbf{E}}^0=-\nabla_{\mathbf{X}}\mathbf{\Phi}^{0}$
    \EndProcedure
    \State $\mathbf{F}=\mathbf{F}^{\text{new}}$ and $\tilde{\mathbf{E}}=\tilde{\mathbf{E}}^{\text{new}}$ \label{energy}
    \State Compute the energy density ${W}(\mathbf{C},\tilde{\mathbf{E}})$ 
    \State Compute stress: $\bar{\mbf{S}}={\partial{W}(\mbf{C},\tilde{\mbf{E}})}/{\partial\mbf{C}}|_{\mbf{C}=\Theta^{2/3}\hat{\mbf{C}}}$
%    \While{$n<n_{end}$}\Comment{We have the answer if r is 0} \label{marker}
    \Procedure{Solve Mechanical}{$\bar{\mbf{S}}$,$\tilde{\mathbf{E}}$}
      \State Compute $\mathbf{f}^{n}_{m}$ (Eq. \ref{eq:ssolids})
      \State Obtain accelerations: $\ddot{\mathbf{u}}^{n}=\mathbf{M}^{-1}\mathbf{f}^{n}_{m}$
      \State $\dot{\mathbf{u}}^{n+\frac{1}{2}}=\dot{\mathbf{u}}^{n-\frac{1}{2}}+\Delta{t}\ddot{\mathbf{u}}^{n}$
      \State $\mathbf{u}^{n+1}=\mathbf{u}^{n}+\Delta{t}\dot{\mathbf{u}}^{n+\frac{1}{2}}$
	  \State \textbf{return} $\mathbf{u}^{n+1}$
  \EndProcedure
  \Procedure{Solve Electrical}{$\mathbf{u}^{n+1}$}
  	  \State Compute $\epsilon J \mathbf{C}^{-1}(\mathbf{u}^{n+1})$
  	  \State Compute $\mathbf{K}^{n+1}_{ee}$ and $\hat{\mathbf{f}}_{e}$ (Eqs. \ref{eq:selectric2} and \ref{eq:selectric})
  	  \State Solving $\mathbf{K}^{n+1}_{ee}\mathbf{\Phi}^{n+1}=\hat{\mathbf{f}}_{e}$
  	  \State \textbf{return} $\mathbf{\Phi}^{n+1}$
  \EndProcedure
  \State Update field variables $\mathbf{u}^{\text{new}}=\mathbf{u}^{n+1}$ and $\mathbf{\Phi}^{\text{new}}=\mathbf{\Phi}^{n+1}$
  \State Update counter $n \leftarrow n+1$
   \State \Goto{energy} if the simulation is not ended
 %     \EndWhile
\end{algorithmic}
\end{algorithm}

\subsection{Discussion on Explicit-Implicit Staggered Formulation}

We now discuss and elaborate upon various aspects of the explicit-implicit staggered formulation.  First, we note that this formulation does not require the calculation of the complex electromechanical coupling stiffnesses $\mbf{K}_{me}=\mbf{K}_{em}^{T}$ as in the  monolithic formulation shown in (\ref{eq:fe2}).  We also note that the right hand side of (\ref{eq:ssolids}) is not a residual as in an implicit-implicit procedure, but the actual difference in external and internal forces.  \hspi{Thus, the proposed staggered explicit-implicit procedure achieves a full second-order accuracy at each integration step without having to perform iterations whereas full-Newton or modified Newton iterations are essential in an implicit-implicit procedure.}

Second, because second variations in the free energy are not required, i.e. only mechanical stresses, and not stiffnesses, are needed, the incompressibility constraint $\lambda$ has a smaller effect on the stable step size in the explicit integration of the structural equations.  Specifically, it was demonstrated in Park and Underwood~\cite{parkCMAME1980} that for nonlinear problems it is the {\it apparent frequency} $\left(\omega_{ap}^{}\right)_{max}$ that dictates the maximum stable integration step size in explicit integration defined as
\begin{align}
\left(\omega^2_{ap}\right)_i &= \frac{\Delta \ubold_i\cdot \Delta \ddot\ubold_i}{\Delta \ubold^2_i},\quad i=1,2,\ldots, \ssi{N_{nd}}. 
\label{eq: apparent frequency}\\
\left(\omega_{ap}\right)_{max} & = \max_{1\leq i \leq \ssi{N_{nd}}} \lbrace \left(\omega_{ap}\right)_i\rbrace
\nonumber
\end{align}
\hspi{where $N_{nd}$ represents the number of FE nodes in the system.}  

Note that the dominant term in the mechanical stiffness operator (see Eq. 27 of Park \emph{et al.}~\cite{parkIJSS2012}) is given by $\lambda C^{-1}_{IJ}C^{-1}_{KJ}$ whereas in the mechanical stress the incompressibility term is manifested in $\lambda \log J  C^{-1}_{IJ}$. To this end, we first identify the $\lambda$-term contributing to $\fbold^{int}$  as follows: 
\begin{gather}
\begin{split}
\Delta \ddot\ubold = \ddot\ubold^{n+1} - \ddot\ubold^n, \quad 
\ddot\ubold & = \Mbold_{uu}^{-1}( \fbold_{ext} - \fbold_{int})
\\
\fbold_{int}   & = \int \bar S_{iJ} \frac{\partial N_a}{\partial X_J} \ dV 
\\
\bar S_{iJ} & = 2 F_{iL} \frac{\partial  W(\Cbold, \tilde\Ebold)}{\partial C_{JL}} 
\\
 2\frac{\partial  W(\Cbold, \tilde\Ebold)}{\partial C_{JL}}
  & =  \lambda \log J \Cbold^{-1}_{JL}   
\\
  & +   2\mu  [W'_{0}(I)  \delta_{JL} - W'_0(3)C^{-1}_{JL}]  
\\
 &+ \epsilon J \tilde E_K\tilde E_I (C^{-1}_{KJ}C^{-1}_{IL} - \half C^{-1}_{KI}C^{-1}_{JL}) 
 \label{eq:fint}
\end{split}
\end{gather}
The contribution to the $i$-th apparent frequency by the incompressible parameter ($\lambda$) is given by
\begin{gather}
\begin{split}
\left[\left(\omega^2_{ap}\right)_i\right]_\lambda & = \frac{\Delta \ubold_i\cdot \mbold_i^{-1} \left[\left(\Delta \fbold_{int}\right)_i\right]_{\lambda}}{\Delta \ubold_{i}^{2}}
\\
\left[\fbold_{int}\right]_\lambda    & = \int  \lambda  \log J\  \gbold(\ubold) \ dV,\quad 
 \gbold(\ubold) =  F_{iL} C^{-1}_{IJ}  \frac{\partial N_a}{\partial X_J} 
\\
& \Downarrow
\\
\left[\Delta \fbold_{int}\right]_\lambda & =  \int  \lambda  \log J^{n+1} \  \gbold (\ubold^{n+1}) \ dV - 
 \int  \lambda  \log J^{n} \  \gbold (\ubold^{n}) \ dV, \quad J \rightarrow 1
\label{eq:partial_internal_force}
\end{split}
\end{gather}
where for brevity in explaining the contribution of the incompressible parameter ($\lambda$), we assumed a diagonal mass matrix ($\mbold_i, i =1,2,\ldots \ssi{N_{nd}}$).
 
Hence, while the incompressible parameter $\lambda$ plays a major role in implicit integration as it is the major material parameter, it plays a minor role in contributing to the apparent frequency magnitude ($\omega_{ap}$) because, as shown in  Eq. (\ref{eq:partial_internal_force}), $\Delta\lambda \log J\rightarrow0$.  This means that the integration step size for explicit integration cases is dominated by the  second term of 
$\frac{\partial \hat W(\Cbold, \tilde\Ebold)}{\partial C_{JL}}$, i.e. the $2\mu  [W'_{0}(I)  \delta_{JL} - W'_0(3)C^{-1}_{JL}]$ term.  

Third, while the equations for the mechanical and electrostatic domains are no longer solved simultaneously as in the monolithic approach in Eq. (\ref{eq:fe2}), the correct coupling effects are accounted for.  This is enabled because the free energy in (\ref{eq:de1}) is electromechanically coupled through the $-\frac{\epsilon}{2}JC_{IJ}^{-1}\tilde{E}_{I}\tilde{E}_{J}$ term.  Therefore, for the mechanical problem, the dielectric contribution to the internal force ($\fbold_{int}$) are accounted for in the third term of $\frac{\partial \hat W(\Cbold, \tilde\Ebold)}{\partial \Cbold_{JL}}$ in Eq. (\ref{eq:fint}) above.  For the electrostatic problem, the stiffness matrix $\mbf{K}_{ee}$ in Eq. (\ref{eq:selectric2}) depends on the structural deformation through the inverse of the stretch tensor $\mbf{C}^{-1}(\mbf{u}^{n+1})$. 

As a final note, one may argue to adopt one of four existing approaches to model the DEs:  a fractional step method~\cite{yanenko1971},  an operator splitting method~\cite{strangSIAM1968}, an adiabatic partial integration algorithm splitting method~\cite{armeroIJNME1992}, and an augmented stabilization method~\cite{farhatCMAME1991}. These methods are appropriate when the evolution of the coupled governing equations is explicitly time-dependent. However, for DEs only the the evolution of the structural system is explicitly time-dependent, whereas the governing electrostatic equations are not explicitly time-dependent.  This fact makes an adoption of fractional step integration a moot point, viz., no advantage is accrued by taking two half-step integration advances to arrive at the full step integration. This is because at each fractional step, the solution vectors are not of intermediate incomplete values but the correct vectors at that time step.   Instead, only the updated displacement $\mbf{u}^{n+1}$ is required to satisfy the solution of the electrostatic equations in (\ref{eq:selectric2}).  This observation plus the simplicity of explicit integration of the structural evolution equation enables the simple staggered approach described above for the analysis of DEs.

\subsection{Stability and accuracy analysis}

It is well known  that the computational stability limit of integrating the structural dynamics equations by the central difference method is given by 
\begin{equation}
\omega_{max}^{} \Delta t \le 2 \
\label{eq: cd-stability}
\end{equation}
where $\omega_{max}^{}$ is the highest discrete frequency of the uncoupled structural dynamical equation, and $\Delta t$ is the integration timestep size.  

Employing the  linearized coupled dynamical form from \eqref{eq:fe2}, one has the following eigenvalue problem:
\begin{gather}
\begin{split}
& \left[\begin{array}{cc}
(\mu \Kbold_{mm} -\omega^2\Mbold) &\epsilon\Kbold_{me}
\\
\epsilon \Kbold_{me}^T  &\epsilon\Kbold_{ee}
\end{array}\right]
 \left\{ \begin{array}{c} \Delta \ubold\\ \Delta\Phibold\end{array} \right\} 
= \left\{ \begin{array}{c}  \zerobold\\  \zerobold\end{array} \right\} 
 \label{eq:coupled-eigenvalue}
\end{split}
\end{gather}
When specialized to a two-degree of freedom model case, one has the following eigenvalue problem:
\begin{gather}
\begin{split}
& \left[\begin{array}{cc}
(\mu \kbold_{mm} -\omega^2\mbold) &\epsilon\kbold_{me}
\\
\epsilon \kbold_{me}^T  &\epsilon\kbold_{ee}
\end{array}\right]
 \left\{ \begin{array}{c} \Delta \ubold\\ \Delta\Phibold\end{array} \right\} 
= \left\{ \begin{array}{c}  \zerobold\\  \zerobold\end{array} \right\} 
 \label{eq:2dof-eigenvalue}
\end{split}
\end{gather}
from which we find  
\begin{equation}
\omega^2_{coupled} = \frac{\mu \kbold_{mm} - \epsilon \kbold_{me}^2 /\kbold_{ee}}{\mbold}\ 
<\  \frac{\mu \kbold_{mm}}{\mbold} = \omega^2_{structure}
 \label{eq:upperbound}
\end{equation}
Eq. (\ref{eq:upperbound}) demonstrates that the frequency of the electromechanically coupled system is smaller than the frequency of the mechanical-only system, which implies that $\Delta t_{structure}<\Delta t_{coupled}$.  Therefore, explicit integration of the electromechanically coupled structural equation by the central difference method employing the step size determined by the highest frequency of the uncoupled structural dynamics equation ensures computational stability.

As for accuracy considerations, the proposed explicit integration of the coupled structural dynamics equations through Eq. (\ref{eq:ssolids}) and implicit solution of the electrostatic equation in Eq. (\ref{eq:selectric}) yields second-order accuracy due to the second-order accuracy properties of the central difference integrator~\cite{hughes1987}.  

The numerical experiments to be discussed in the following section will serve to corroborate the computational stability, accuracy assessments as well as the justification for the explicit-implicit staggered methodology discussed in the present section.

\section{Numerical Examples}

We now present 2D and 3D numerical examples verifying the accuracy and efficiency of the proposed staggered methodology as compared to previously developed monolithic approaches~\cite{parkIJSS2012,parkCMAME2013,seifiIJSS2016} for electroactive polymers.  The staggered explicit-implicit formulation was implemented into the open source simulation code Tahoe~\cite{tahoe}, which was previously where the monolithic approach was implemented.  All examples involve electromechanical instabilities, i.e. wrinkling, creasing and bursting drops, to demonstrate the robustness of the proposed approach.  

\subsection{Surface Tension-Driven Creasing to Wrinkling Transition in a 2D Film}

\begin{figure} 
\centering
\includegraphics[scale=0.75]{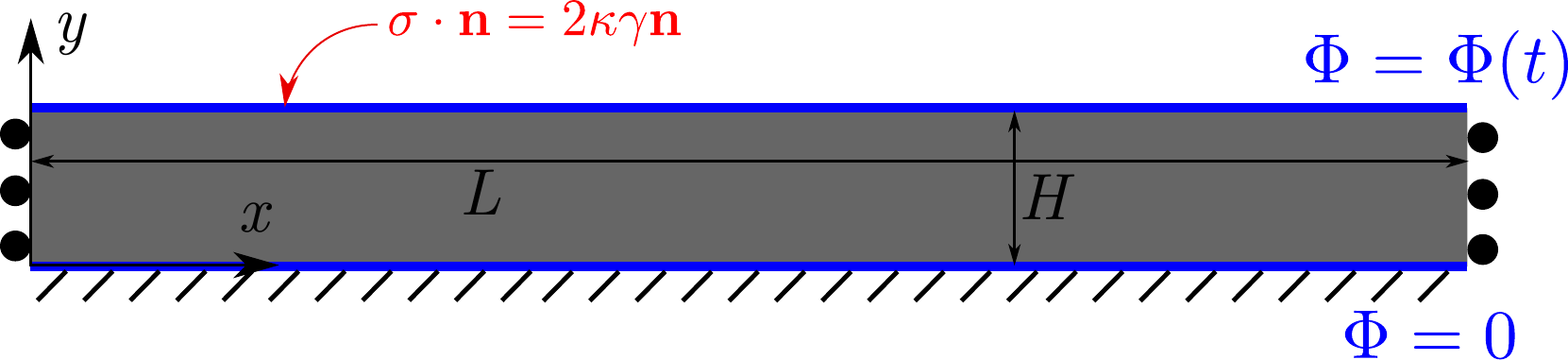}
\caption{Schematic of the computational model showing a 2D film with electro-elasto-capillary boundary conditions based on experiment by Wang \emph{et al.}~\cite{wangPRE2013}.}
\label{fig:film_schematic}
\end{figure}

Our first numerical example considers a 2D, plane strain DE film as shown in Figure (\ref{fig:film_schematic}).  Previous experiments~\cite{wangPRE2013}, and numerical simulations~\cite{seifiIJSS2016} have demonstrated that as the surface tension $\gamma$ on the top surface increases, the electromechanical surface instability that occurs transitions from creasing to wrinkling.  This instability transition is driven by the surface tension driving force reduced surface area, leading to a smoother, longer wavelength surface instability.  

The film was fixed mechanically at the bottom surface ($y=0$), with rollers on both the left and right sides. The electrostatic boundary conditions were that the voltage on the bottom surface was kept at zero, i.e. $\Phi=0$, while the voltage on the top surface was subject to a linearly increasing voltage with time, i.e. $\Phi=\Phi(t)$.  Additionally in order to account for elastocapillary effects, the top surface was also subject to the Young-Laplace equations $\sigma\cdot\mathbf{n}=2 \kappa\gamma\mathbf{n}$, where $\kappa$ is the mean curvature, $\gamma$ is the surface tension and $\mathbf{n}$ is the normal vector to the surface.  The elastocapillary force resulting from the surface tension augments the right hand side of (\ref{eq:ssolids}) as 
\begin{equation}
  \fbold_s=\int \gamma \nabla_{s}\mbf{N}da
\end{equation}
where $\nabla_{s}=(\mbf{I}-\mbf{n}\otimes\mbf{n})\nabla$ is the surface gradient operator.  The dimensions of the film were $L=160$ and $H=4$, where the film was discretized with standard 4-node bilinear quadrilateral finite elements.  For both the staggered and monolithic solutions, 640 4-node elements were utilized, while a time step of $\Delta t=0.01$ was chosen for both models.  \hspi{The same time step was chosen for both models so that we could, as close as possible, provide an apples to apples comparison with regards to the computational expense of the staggered and monolithic formulations}.  For this and all subsequent examples, the Q1P0 approach of Simo \emph{et al.}~\cite{simoCMAME1985} was used for both the staggered and monolithic methods to mitigate the effects of volumetric locking.

\begin{figure}
\centering
\includegraphics[scale=0.5]{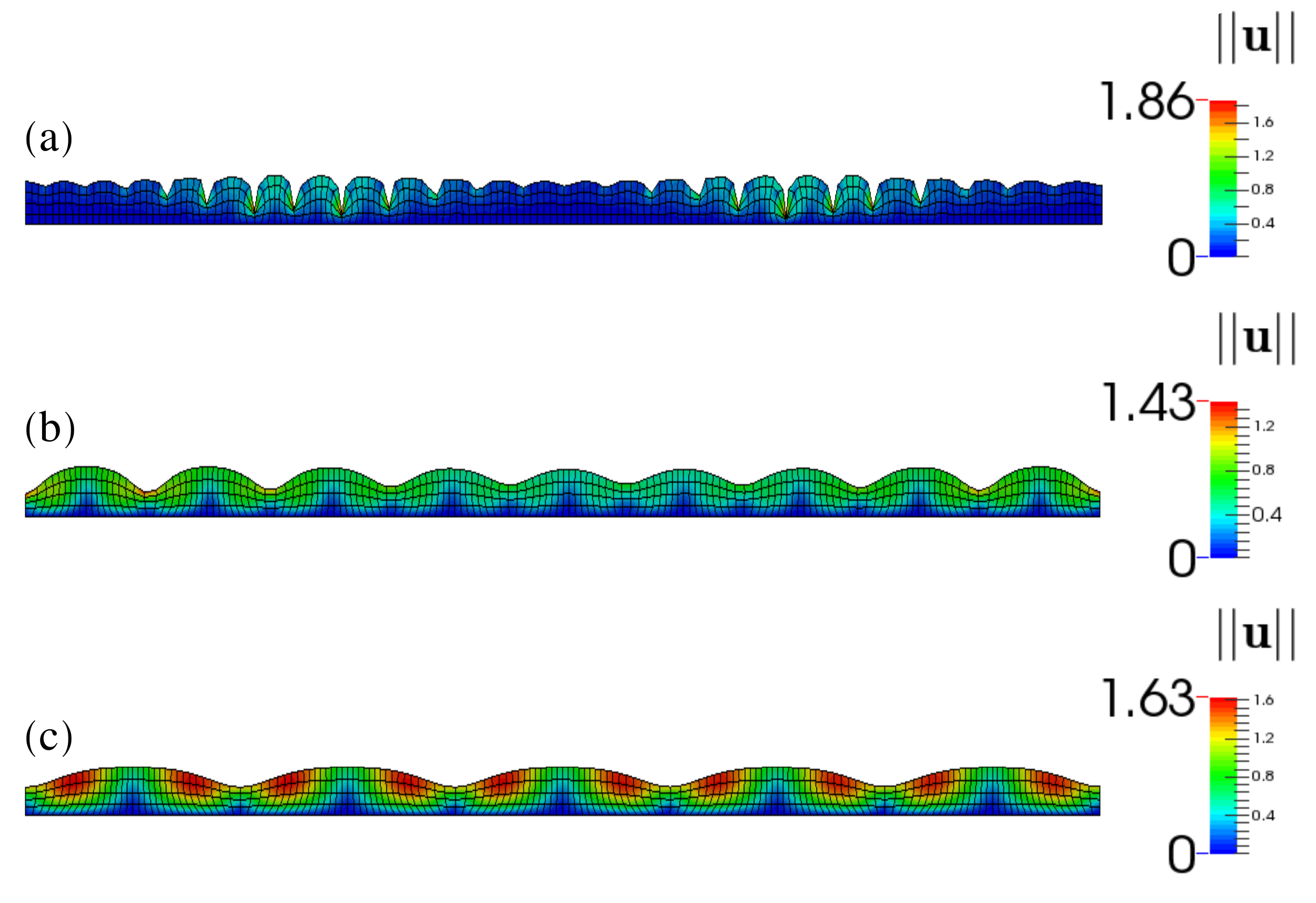}
\caption{Creasing to wrinkling transition using staggered explicit-implicit method for a DE film with dimensions $L=160$ and $H=4$ for three different elasto-capillary numbers $\bar{\gamma}=\gamma/(\mu H)$:  (a) $\bar{\gamma}=0.25$;  (b) $\bar{\gamma}=2$; (c) $\bar{\gamma}=16$. $||\mathbf{u}||$ denotes the displacement magnitude.}
\label{fig:film_explicit}
\end{figure}

\begin{figure}
\centering
\includegraphics[scale=0.5]{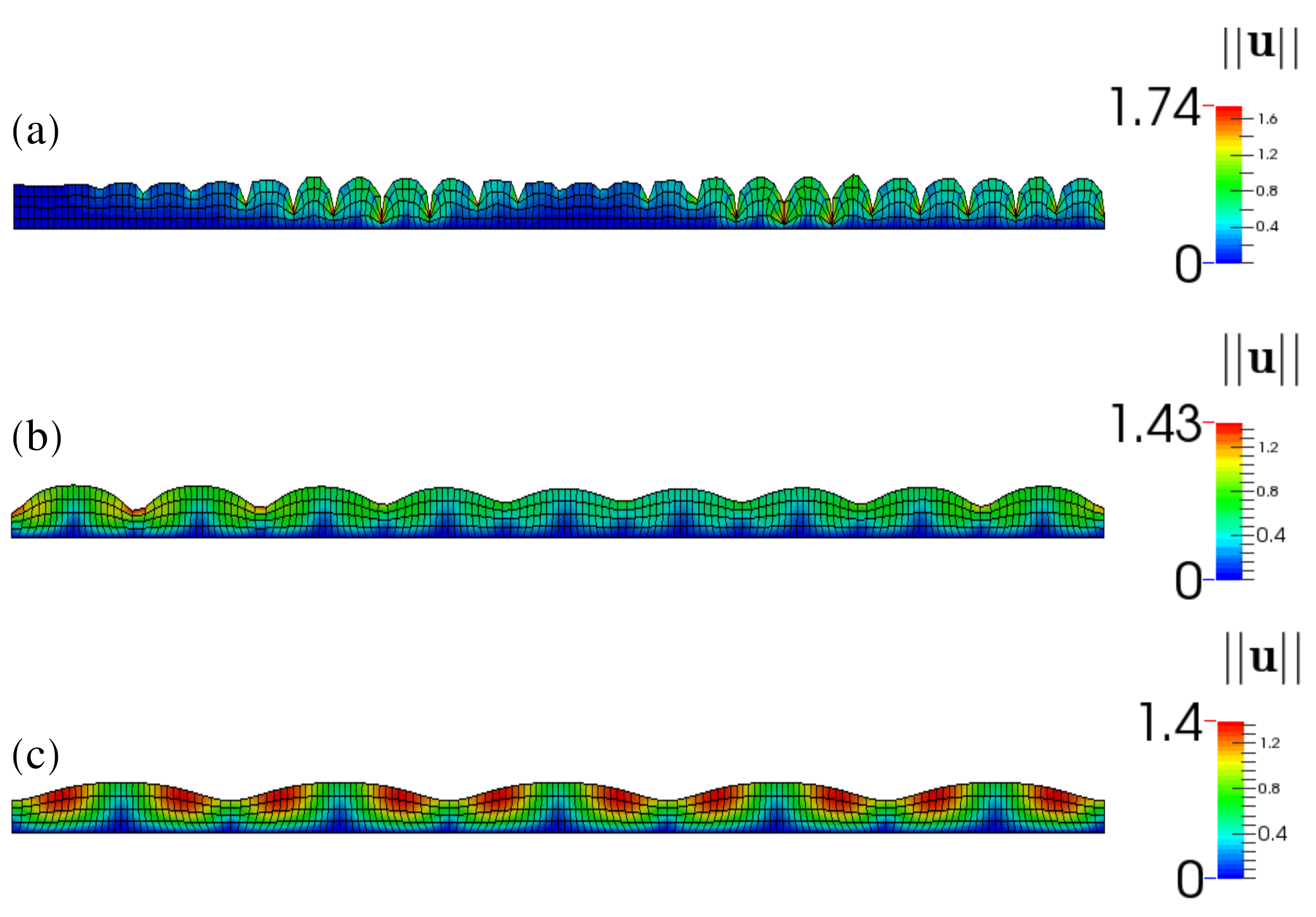}
\caption{Creasing to wrinkling transition using fully coupled, monolithic method for a DE film with dimensions $L=160$ and $H=4$ for three different elasto-capillary numbers $\bar{\gamma}=\gamma/(\mu H)$:  (a) $\bar{\gamma}=0.25$;  (b) $\bar{\gamma}=2$; (c) $\bar{\gamma}=16$. $||\mathbf{u}||$ denotes the displacement magnitude.}
\label{fig:film_fully}
\end{figure}

Figures (\ref{fig:film_explicit}) and (\ref{fig:film_fully}) show the surface creasing to wrinkling transition for the staggered explicit-implicit and monolithic methods, respectively.  In both Figures (\ref{fig:film_explicit})(a) and (\ref{fig:film_fully})(a), a short wavelength surface creasing instability is observed for elastocapillary numbers $\bar{\gamma}=\gamma/(\mu H)$ that are smaller than unity, where $\mu$ is the shear modulus.  As the elastocapillary number increases beyond unity in Figures (\ref{fig:film_explicit})(b-c) and (\ref{fig:film_fully})(b-c), a transition to a smoother, longer wavelength wrinkling instability is observed, where the wrinkling wavelength increases with increasing elastocapillary number.  A comparison between Figures (\ref{fig:film_explicit}) and (\ref{fig:film_fully}) demonstrates the similarity between the staggered and monolithic solutions.  Furthermore, the creasing to wrinkling transition shown here is consistent with previous experimental~\cite{wangPRE2013} and computational~\cite{seifiIJSS2016} studies.  

%\begin{table}[]
%\centering
%\caption{Computation time for 2D case with 640 elements (\sss{this is temporarily, I'll probably do plotting instead of table})}
%\label{my-label}
%\begin{tabular}{l || l | l}
%           & explicit-implicit & fully coupled   \\
%$\gamma_1$ & 6737 sec          & 39949 sec       \\
%$\gamma_2$ & 9754 sec          & 62934 sec       \\
%$\gamma_3$ & 12765 sec                  & 92552 sec      
%\end{tabular}
%\end{table}

%\begin{figure}
%\centering
%\includegraphics[scale=0.75]{pics/Ec.pdf}
%\caption{Elasto-capillary number $\gamma/(\mu H)$ vs. critical electric field $E_c\sqrt{\epsilon/\mu}$ for both monolithic and staggered explicit-implicit schemes. \hsp{Please change captions to monolithic and staggered}}
%\label{fig:Ec}
%\end{figure}

\begin{figure}
\centering
\begin{tikzpicture} 
\begin{axis}[xmode=log,		
		xmin=0.001,
		ymin=0.75,
		xmax=100, xlabel=$\gamma/(\mu H)$, ylabel=$E_c\sqrt{\epsilon/\mu}$,
		legend pos=north west,	
		legend style={draw=none},	
		legend entries={monolithic,staggered}] 
\addplot[color=blue, dash pattern=on 5pt, thick] coordinates { 
(0.00, 1.267) 
(0.01, 1.395) 
(0.0625, 1.485) 
(0.125, 1.69) 
(0.25, 1.97) 
(0.5, 2.185) 
(1.0, 2.495) 
(2.0, 2.9)
(4.0, 3.4)
(8.0, 3.94) 
(16.0, 4.56)};
\addplot[color=black,only marks, mark=x, mark size=4pt] coordinates { 
(0.00, 1.245) 
(0.01, 1.42) 
(0.0625, 1.535) 
(0.125, 1.695) 
(0.25, 1.975) 
(0.5, 2.197) 
(1.0, 2.490) 
(2.0, 2.90)
(4.0, 3.40)
(8.0, 3.95) 
(16.0, 4.55)}; 
\end{axis} 
\end{tikzpicture}
\caption{Elasto-capillary number $\gamma/(\mu H)$ vs. critical electric field $E_c\sqrt{\epsilon/\mu}$ for both monolithic and staggered explicit-implicit schemes.}
\label{fig:Ec}
\end{figure}
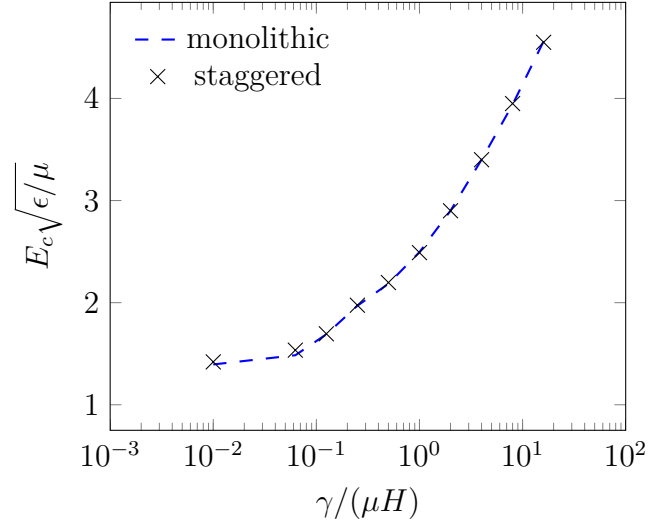

%\begin{figure}
%\centering
%\includegraphics[scale=0.75]{pics/W.pdf}
%\caption{Elasto-capillary number vs. wavelength $\ell$ for both monolithic and staggered explicit-implicit schemes.  \hsp{Please change captions to monolithic and staggered.  On the y-axis, please add $\lambda$ as the wavelength, i.e. $\ell=l/H$}\sss{changed $\lambda$ to $\ell$ in order to avoid having conflicting notatio (bulk modulus is $\lambda$) }}
%\label{fig:W}
%\end{figure}

\begin{figure}
\centering
\begin{tikzpicture} 
\begin{axis}[xmode=log,		
		xmin=0.001,
		ymin=0.75,
		xmax=100, xlabel=$\gamma/(\mu H)$, ylabel=$\ell/H$,
		legend pos=north west,	
		legend style={draw=none},	
		legend entries={monolithic,staggered}] 
\addplot[color=blue, dash pattern=on 5pt, thick] coordinates { 
(0.00, 1.5) 
(0.01, 1.53) 
(0.0625, 1.6) 
(0.125, 1.65) 
(0.25, 1.715) 
(0.5, 2.037) 
(1.0, 2.593) 
(2.0, 3.337)
(4.0, 4.44)
(8.0, 5.714) 
(16.0, 8)};
\addplot[color=black,only marks, mark=x, mark size=4pt] coordinates { 
(0.00, 1.5) 
(0.01, 1.53) 
(0.0625, 1.66) 
(0.125, 1.73) 
(0.25, 1.739) 
(0.5, 2.0) 
(1.0, 2.54) 
(2.0, 3.33)
(4.0, 4.44)
(8.0, 5.71) 
(16.0, 8)}; 
\end{axis} 
\end{tikzpicture}
\caption{Elasto-capillary number vs. wavelength $\ell/H$ for both monolithic and staggered explicit-implicit schemes. }
\label{fig:W}
\end{figure}
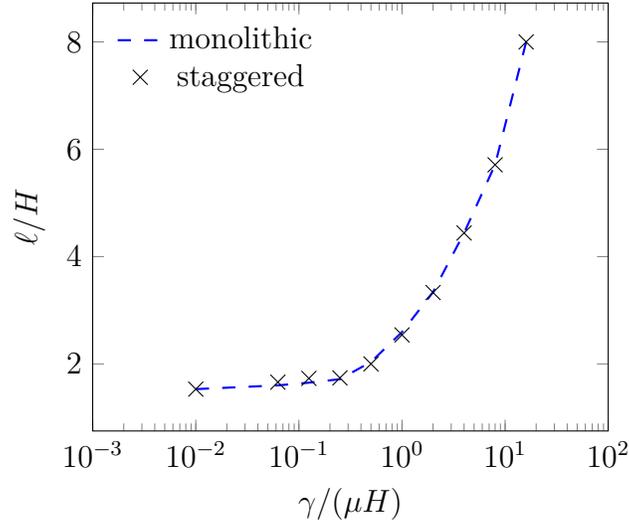

To provide a more precise comparison between the monolithic and staggered results, we also measured the normalized critical electric field $E_c\sqrt{\epsilon/\mu}$ at the onset of surface instability as a function of the elasto-capillary number $\gamma/(\mu H)$ for both the monolithic and staggered models, where $\epsilon$ is the dielectric constant. The results show excellent agreement with one another, as shown in Figure (\ref{fig:Ec}).  Finally, we measured the wavelength $\ell=l/H$ of the creases and wrinkles formed on the surface, where $l$ is the distance between creases or wrinkles, and $H$ is the film thickness.  Figure (\ref{fig:W}) demonstrates that there is excellent agreement on the wavelength as a function of elastocapillary number between the monolithic and staggered approaches, where the accuracy of the monolithic model was previously shown in the work of Seifi and Park~\cite{seifiIJSS2016}.

%\begin{figure}
%\centering
%\includegraphics[scale=0.75]{pics/crease_time.pdf}
%\caption{\sss{Elapsed time vs. number of elements ($n_e$) of simple 2D plane-strain crease problem for a DE with $H\times L=4\times 20$ with mesh spacings $h=2,\ 1,\ 0.5 \text{ and }0.25$ and time step $\Delta t=0.005$. The elapsed time is at the onset of instability.}}
%\label{fig:crease_time}
%\end{figure}

%\begin{figure}
%\centering
%\begin{tikzpicture} 
%\begin{axis}[xmode=log,		
%		xmin=10, xmax=10000, xlabel=$n_{dof}$, ylabel=$t_m/t_s$] 
%\addplot[color=blue, mark=x, mark size=4pt, thick] coordinates { 
%(66, 11.6897) 
%(210, 13.4401) 
%(738, 16.8341) 
%(2754, 20.5684) 
%};
%\end{axis} 
%\end{tikzpicture}
%\caption{\sss{Ratio of elapsed time for monolithic model over elapsed time for staggered model ($t_m/t_s$) as a function of total numbers of degree of freedoms $n_{dof}$. }}
%\label{fig:crease_time}
%\end{figure}

%\begin{figure}
%\centering
%\includegraphics[scale=0.75]{pics/crease_ratiotime.pdf}
%\caption{\sss{Ratio of elapsed time for monolithic model over elapsed time for staggered model ($t_m/t_s$) as a function of total numbers of degree of freedoms $n_{dof}$. The elapsed time calculated at onset of crease. }}
%\label{fig:crease_time}
%\end{figure}

\subsection{Bursting Drops in a 2D Plane Strain Film}

\begin{figure} 
\centering
\includegraphics[scale=1]{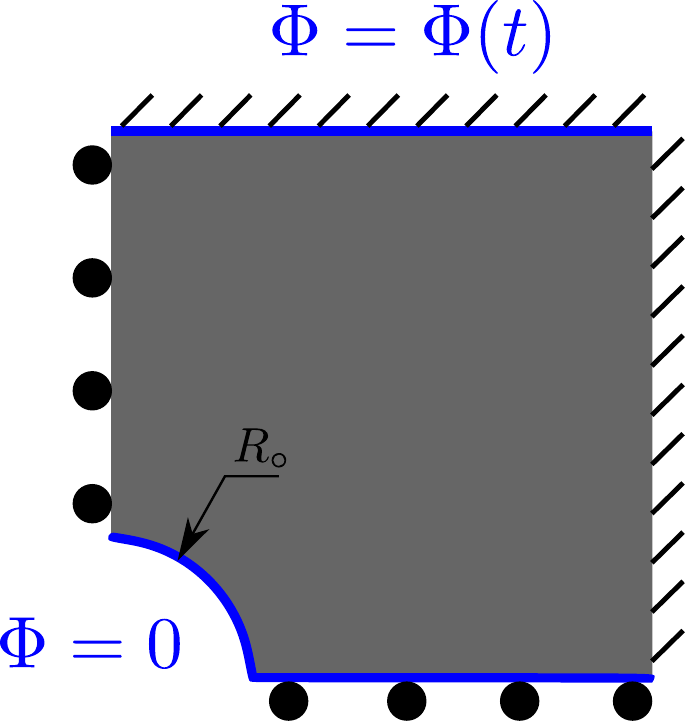}
\caption{Axisymmetric computational model for
bursting drop in a dielectric solid based on experiment by Wang et al. \cite{wangNC2012}.}
\label{fig:droplet_schematic}
\end{figure}

Our second numerical example in 2D considers the case of a bursting drop, as shown in Figure (\ref{fig:droplet_schematic}).  In this problem, the electromechanical instability of interest revolves around a small droplet of conductive fluid contained within a DE, which elongates in a crack-like fashion towards the boundaries of the DE where the voltage is applied.  This example has also been studied experimentally~\cite{wangNC2012}, and computationally~\cite{parkSM2013,seifiIJSS2016}. 

We performed numerical simulations using both the monolithic and staggered models by utilizing the one quarter computational domain with the electromechanical boundary conditions shown in Figure (\ref{fig:droplet_schematic}). This model had dimensions $20\times 20$ with the radius of the quarter circular hole being $R_{\circ}=2$. This axisymmetric domain was again discretized using standard 4-node bilinear quadrilateral finite elements with a mesh size of unity. The voltage was prescribed to be zero along the hole perimeter and along the bottom surface, while the top surface was subject to a voltage that increased linearly in time.  

\begin{figure}
\centering
\includegraphics[scale=0.25]{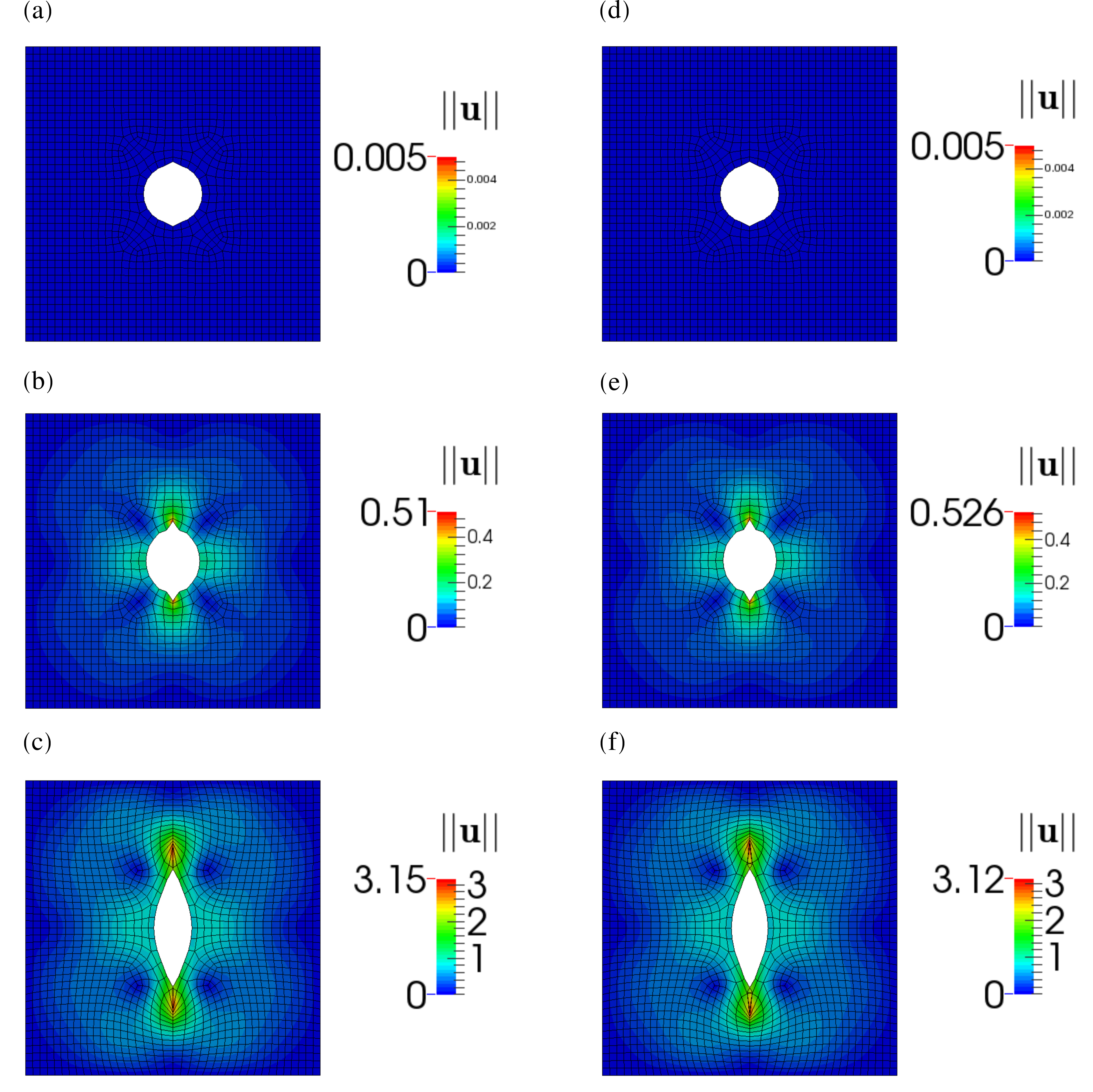}
\caption{Fully coupled, monolithic solution on the left side (a)-(c) vs. staggered, explicit-implicit solution on the right side (d)-(f) for the bursting drop problem. (a) and (d) at $t=0$, (b) and (e) at $t=t_{mid}$ and (c) and (f) at $t=t_{final}$.  $||\mathbf{u}||$ denotes the displacement magnitude.}
\label{fig:drop_explicit_fully}
\end{figure}

The time evolution of the drop for both the monolithic and staggered methods is shown in Figure (\ref{fig:drop_explicit_fully}). The figure shows three stages of deformation of the droplet subjected to the applied electric field, starting at the point where the drop elongation has just begun in Figure (\ref{fig:drop_explicit_fully})(a) and (d), along with two other comparisons between the monolithic and staggered formulations in Figures (\ref{fig:drop_explicit_fully})(b) and (e) and also Figures (\ref{fig:drop_explicit_fully})(c) and (f).  In all cases, the drop configuration compares well between the monolithic and staggered solutions.  

%\begin{figure}
%\centering
%\includegraphics[scale=0.75]{pics/droplet.pdf}
%\caption{Normalized position of the bursting drop tip $b/R_{\circ}$ as a function of applied electric field $E_{top}$, where $R_{\circ}$ is the initial radius of the drop, and $b$ is the long axis of the bursting drop.  \hsp{As before, change figure legend to monolithic and staggered.}}
%\label{fig:crack_distance}  
%\end{figure}

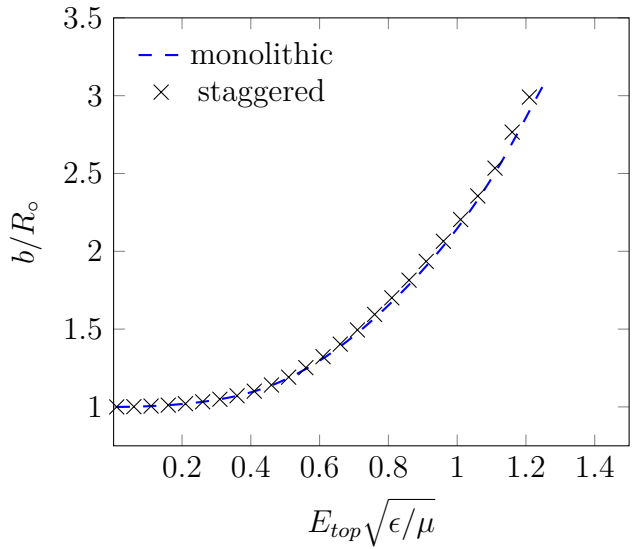
\begin{figure}
\centering
\begin{tikzpicture} 
\begin{axis}[		
		xmin=0.001,
		ymax=3.5,
		xmax=1.5, xlabel=$E_{top}\sqrt{\epsilon/\mu}$, ylabel=$b/R_{\circ}$,
		legend pos=north west,	
		legend style={draw=none},	
		legend entries={monolithic,staggered}] 
\addplot[color=blue, dash pattern=on 5pt, thick] coordinates { 
(	0.01	,	1	)
(	0.02	,	1.0001	)
(	0.03	,	1.0004	)
(	0.04	,	1.0006	)
(	0.05	,	1.001	)
(	0.06	,	1.0015	)
(	0.07	,	1.0021	)
(	0.08	,	1.0028	)
(	0.09	,	1.0035	)
(	0.1	,	1.0044	)
(	0.11	,	1.0053	)
(	0.12	,	1.0064	)
(	0.13	,	1.0075	)
(	0.14	,	1.0087	)
(	0.15	,	1.0101	)
(	0.16	,	1.0115	)
(	0.17	,	1.0131	)
(	0.18	,	1.0148	)
(	0.19	,	1.0165	)
(	0.2	,	1.0185	)
(	0.21	,	1.0205	)
(	0.22	,	1.0227	)
(	0.23	,	1.025	)
(	0.24	,	1.0275	)
(	0.25	,	1.03	)
(	0.26	,	1.0328	)
(	0.27	,	1.0357	)
(	0.28	,	1.0388	)
(	0.29	,	1.0421	)
(	0.3	,	1.0456	)
(	0.31	,	1.0493	)
(	0.32	,	1.0531	)
(	0.33	,	1.0573	)
(	0.34	,	1.0616	)
(	0.35	,	1.0662	)
(	0.36	,	1.0711	)
(	0.37	,	1.0763	)
(	0.38	,	1.0818	)
(	0.39	,	1.0876	)
(	0.4	,	1.0937	)
(	0.41	,	1.1002	)
(	0.42	,	1.107	)
(	0.43	,	1.1142	)
(	0.44	,	1.1219	)
(	0.45	,	1.1298	)
(	0.46	,	1.1382	)
(	0.47	,	1.147	)
(	0.48	,	1.1562	)
(	0.49	,	1.1658	)
(	0.5	,	1.1758	)
(	0.51	,	1.1862	)
(	0.52	,	1.197	)
(	0.53	,	1.2082	)
(	0.54	,	1.2197	)
(	0.55	,	1.2317	)
(	0.56	,	1.244	)
(	0.57	,	1.2567	)
(	0.58	,	1.2698	)
(	0.59	,	1.2832	)
(	0.6	,	1.2971	)
(	0.61	,	1.3113	)
(	0.62	,	1.3259	)
(	0.63	,	1.341	)
(	0.64	,	1.3564	)
(	0.65	,	1.3722	)
(	0.66	,	1.3884	)
(	0.67	,	1.405	)
(	0.68	,	1.422	)
(	0.69	,	1.4393	)
(	0.7	,	1.4571	)
(	0.71	,	1.4751	)
(	0.72	,	1.4936	)
(	0.73	,	1.5123	)
(	0.74	,	1.5314	)
(	0.75	,	1.5508	)
(	0.76	,	1.5706	)
(	0.77	,	1.5907	)
(	0.78	,	1.611	)
(	0.79	,	1.6317	)
(	0.8	,	1.6527	)
(	0.81	,	1.6739	)
(	0.82	,	1.6955	)
(	0.83	,	1.7174	)
(	0.84	,	1.7396	)
(	0.85	,	1.7621	)
(	0.86	,	1.785	)
(	0.87	,	1.8082	)
(	0.88	,	1.8317	)
(	0.89	,	1.8556	)
(	0.9	,	1.88	)
(	0.91	,	1.9046	)
(	0.92	,	1.9297	)
(	0.93	,	1.9552	)
(	0.94	,	1.9811	)
(	0.95	,	2.0075	)
(	0.96	,	2.0343	)
(	0.97	,	2.0615	)
(	0.98	,	2.0894	)
(	0.99	,	2.1176	)
(	1	,	2.1464	)
(	1.01	,	2.1759	)
(	1.02	,	2.2058	)
(	1.03	,	2.2364	)
(	1.04	,	2.2675	)
(	1.05	,	2.2993	)
(	1.06	,	2.3317	)
(	1.07	,	2.3648	)
(	1.08	,	2.3987	)
(	1.09	,	2.4333	)
(	1.1	,	2.4685	)
(	1.11	,	2.5045	)
(	1.12	,	2.5413	)
(	1.13	,	2.5788	)
(	1.14	,	2.6171	)
(	1.15	,	2.656	)
(	1.16	,	2.6957	)
(	1.17	,	2.7359	)
(	1.18	,	2.7767	)
(	1.19	,	2.8179	)
(	1.2	,	2.8594	)
(	1.21	,	2.9011	)
(	1.22	,	2.9427	)
(	1.23	,	2.984	)
(	1.24	,	3.0249	)
(	1.25	,	3.065	)
(	1.26	,	3.1039	)
};
\addplot[color=black,only marks, mark=x, mark size=4pt] coordinates { 
(	0.01	,	1	)
(	0.06	,	1.0015	)
(	0.11	,	1.0053	)
(	0.16	,	1.0115	)
(	0.21	,	1.0205	)
(	0.26	,	1.0329	)
(	0.31	,	1.0494	)
(	0.36	,	1.0715	)
(	0.41	,	1.1012	)
(	0.46	,	1.1407	)
(	0.51	,	1.1912	)
(	0.56	,	1.2521	)
(	0.61	,	1.3228	)
(	0.66	,	1.4035	)
(	0.71	,	1.4945	)
(	0.76	,	1.5943	)
(	0.81	,	1.7011	)
(	0.86	,	1.8145	)
(	0.91	,	1.9352	)
(	0.96	,	2.0646	)
(	1.01	,	2.2041	)
(	1.06	,	2.3555	)
(	1.11	,	2.5345	)
(	1.16	,	2.7657	)
(	1.21	,	2.9911	)
};
\end{axis} 
\end{tikzpicture}
\caption{Normalized position of the bursting drop tip $b/R_{\circ}$ as a function of applied electric field $E_{top}$, where $R_{\circ}$ is the initial radius of the drop, and $b$ is the long axis of the bursting drop.}
\label{fig:crack_distance} 
\end{figure}

Besides the pictorial comparison of the time evolution of the bursting drop configuration in Figure (\ref{fig:drop_explicit_fully}), we also plot the position of the bursting drop tip as a function of applied electric field in Figure (\ref{fig:crack_distance}), which demonstrates that the position of the bursting drop tip as a function of the applied electric field is captured nearly identically between the monolithic and staggered methods.

\subsection{3D Example}

\begin{figure}
\centering
\includegraphics[scale=0.3]{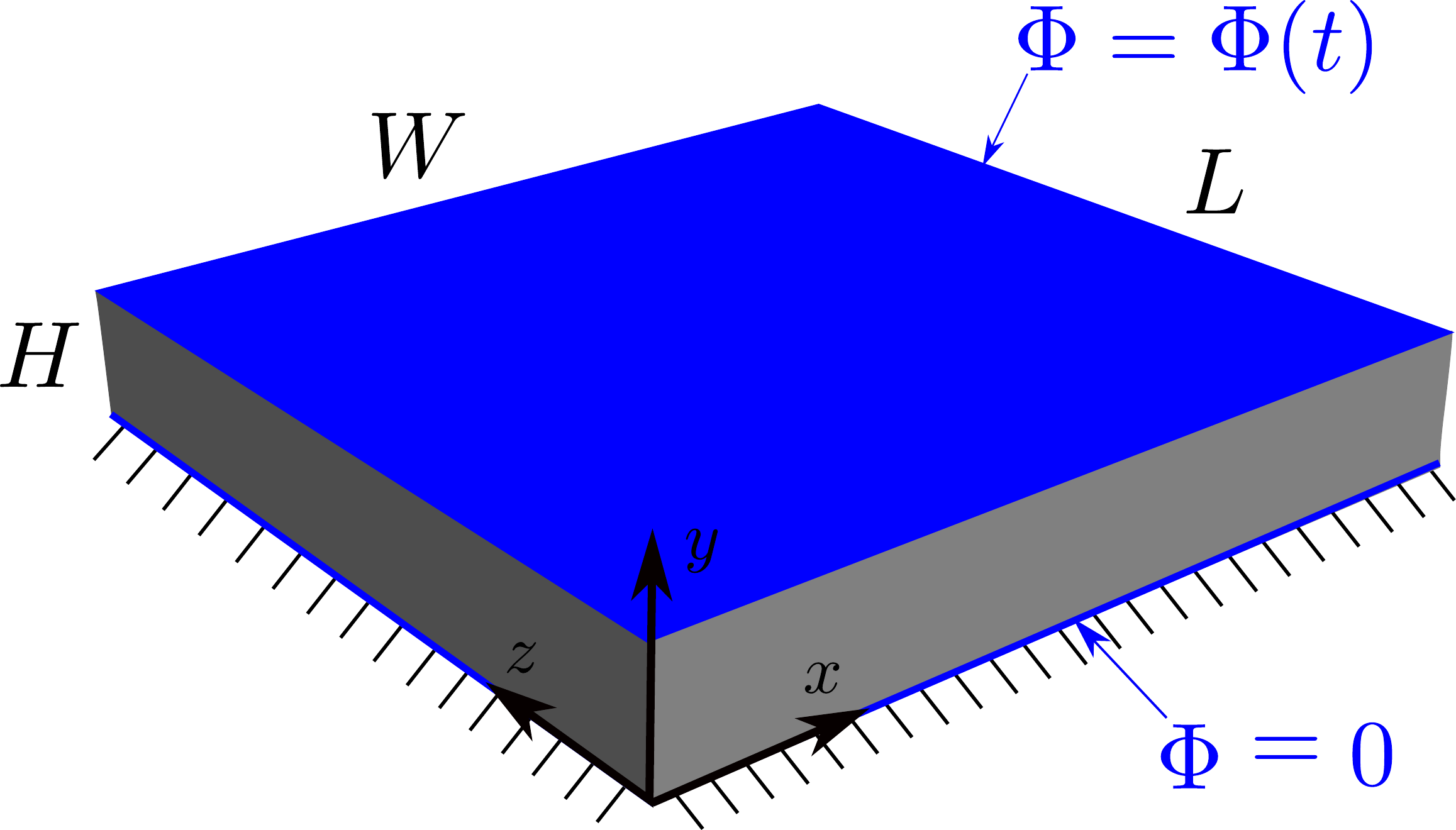}
\caption{3D computational model for creasing of DE plate with dimensions $H\times W\times L$.  The plate is fixed at the bottom with rollers on all sides, while a monolithically increasing voltage $\Phi=\Phi(t)$ is applied to the top surface while the bottom surface remains voltage-free.}
\label{fig:3D_schematic}
\end{figure}

In our final example, we demonstrate the computational efficiency of the staggered methodology by examining a problem involving creasing electromechanical instability in 3D.  Some previous studies have considered 3D problems~\cite{vuIJNME2007,schloglCMAME2016}, but only for simple geometries without complex electromechanical instabilities.  

\begin{figure}
\centering
\includegraphics[scale=0.4]{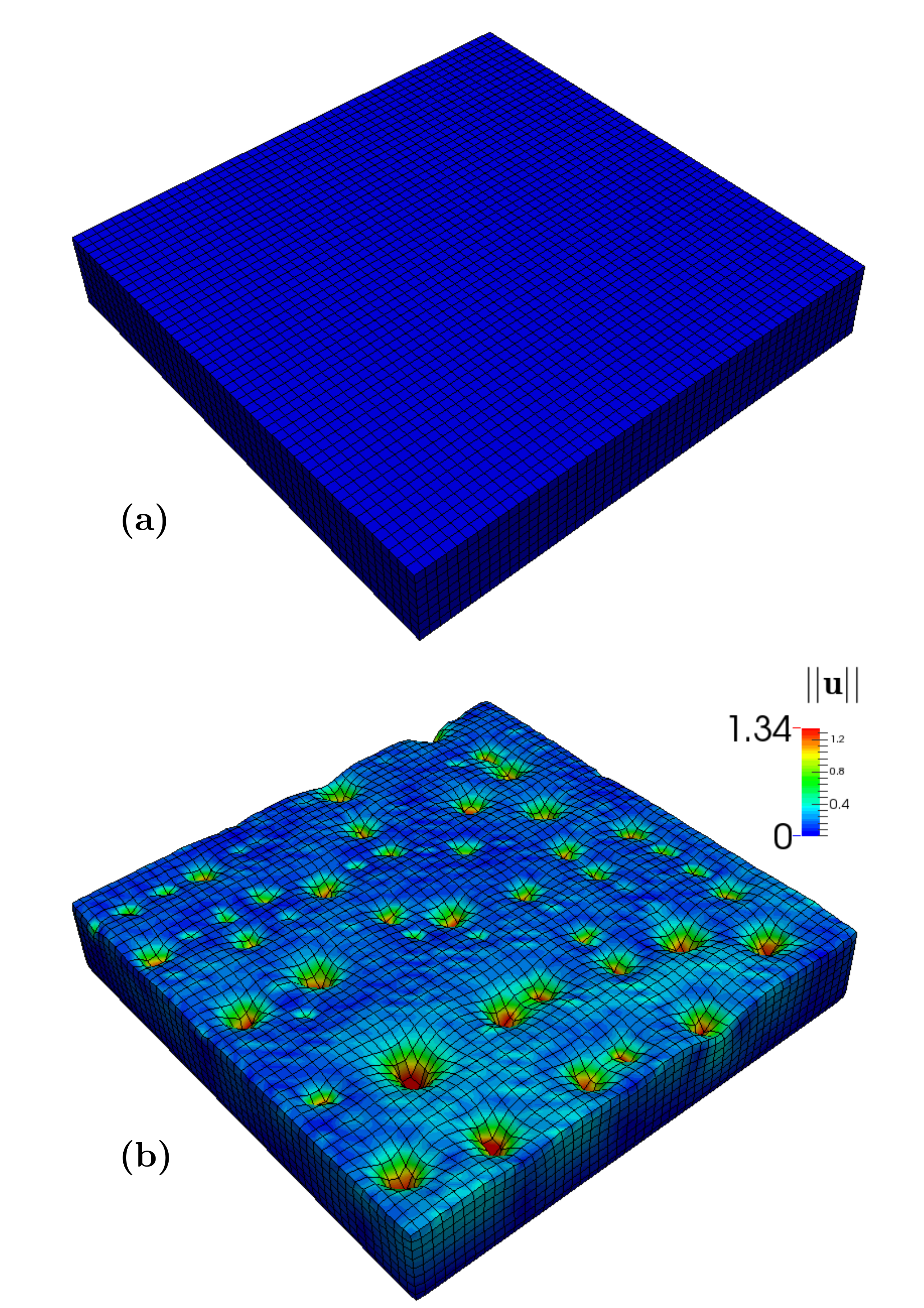}
\caption{3D simulations of creasing of a DE film.  (a) initial undeformed configuration; (b) the deformed configuration shows the creased surface with normalized wavelength $\ell=l/H\approx 1.46$ where the critical electric field is $E_c\sqrt{\epsilon/\mu}\approx 1.08$. }
\label{fig:3Dcrease}
\end{figure}

Here, we modeled the problem involving creasing of a 3D DE film that we showed in the first numerical example through the 2D, plane strain approximation.  The computational domain with dimensions $H\times L\times W=4\times 25 \times 25$ shown in Figure (\ref{fig:3Dcrease})(a) is modeled using standard 8-node hexahedral finite elements with a mesh spacing of $0.5$, giving 20000 finite elements in total, while the same time step of $\Delta t=0.005$ was used for both the staggered and monolithic solutions. The boundary conditions are an extension of the 2D problem as the bottom surface is fixed, while all transverse surfaces are on rollers.  The electrostatic boundary conditions is specified similar to the 2D problem, i.e. with a zero voltage prescribed on the bottom surface while the top surface is subject to a voltage that linearly increases with time.

The result of this simulation is shown in Figure (\ref{fig:3Dcrease})(b). The surface creasing instability occurs when the electric field is $E_c=1.08$, which is in good agreement with previous theoretical predictions~\cite{wangPRL2011a}.   Furthermore, we found that the creasing wavelength is about $\ell=l/H\approx 1.46$ which is quite close to the creasing wavelength of $\ell=l/H\approx 1.5$ found for the 2D problem.  For both the 2D and 3D problems, the creasing wavelengths found are very close to the experimental and analytic solution of $\ell=l/H=1.5$~\cite{wangPRL2011a}, demonstrating the accuracy of the staggered formulation.

\begin{figure}
\centering
\begin{tikzpicture} 
\begin{axis}[xmode=log,		
		xmin=1, xmax=100000, xlabel=$n_{dof}$, ylabel=$t_m/t_s$,
		legend style={draw=none},
		legend pos=north west,	
		legend entries={3D}] 
\addplot[color=blue, mark=x, mark size=4pt, thick] coordinates { 
(15, 2.933346206) 
(135, 10.011944815) 
(1071, 15.72087866) 
(8415, 51.190539984) 
};
%\addplot[color=red, mark=o, mark size=4pt, thick] coordinates { 
%(40, 5.55869) 
%(223, 6.8301) 
%(927, 8.5797101) 
%(3775, 20.5684)};
\end{axis} 
\end{tikzpicture}
\caption{Ratio of elapsed time for monolithic model over elapsed time for staggered model ($t_m/t_s$) as a function of the total numbers of unconstrained degrees of freedoms $n_{dof}$ for the 3D creasing problem.}
\label{fig:3D_time}
\end{figure}
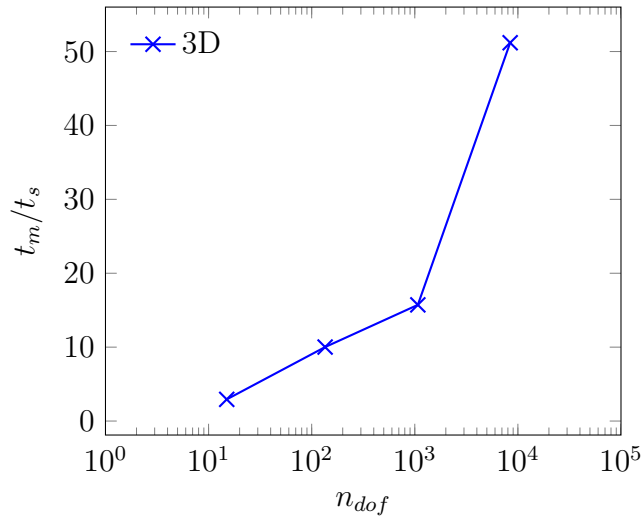

We finally discuss the benefits in computational expense reduction that may be gained through utilization of the staggered approach.  Specifically, we show in Figure (\ref{fig:3D_time}) a comparison of the normalized computational time $t_{m}/t_{s}$, where $t_{m}$ represents the total simulation time for the monolithic approach, and where $t_{s}$ represents the total simulation time for the staggered approach, with both numbers taken for different mesh sizes for the 3D creasing problem.  As expected, there is a significant decrease in computational expense for the staggered method, particularly when the number of degrees of freedom exceeds about 1000.

\section{Conclusions}

%In conclusion, we have proposed a staggered, explicit-implicit formulation for systems, such as electroactive polymers, that are governed by a coupling between Gauss's law for electrostatics and the momentum equation for the mechanical domain.  We demonstrated that the staggered approach is made possible due to the separation in the magnitudes of the mechanical stiffness matrix and the electromechanical coupling matrices, which implies that the solution of the structural problem can be decoupled from the solution of the electrostatic problem, thus enabling an explicit solution of the structural equations.  In contrast, the solution of the structural problem strongly couples to the electrostatic problem, leading to the solution of the electrostatic problem in a nonlinear fashion.  We note that while we did not perform any parallel computations, the explicit solution of the structural problem opens up standard parallel computing capabilities that can be used to solve larger problems with significantly more mechanical and electrostatic degrees of freedom.

In conclusion, we have provided theoretical justification for the stability and accuracy of a simple staggered, explicit-implicit finite element formulation for systems, such as electroactive polymers, that are governed by a coupling between Gauss's law for electrostatics and the momentum equation for the mechanical domain.  The full electromechanical coupling is enabled through the free energy, which enables the correct coupling to enter into both the finite element-discretized momentum and electrostatic equations.

The staggered formulation was shown to give identical solutions to the monolithic formulation for a range of problems involving electromechanical instabilities, though obviously at a significant reduction in computational expense.  While the monolithic formulation has enabled significant insights into the electromechanics of dielectric elastomers for 2D, plane strain problems~\cite{parkCMAME2013,parkSM2013,seifiIJSS2016,seifiSM2017}, very few studies on such instabilities have been performed in 3D.  We anticipate this is where the presently proposed staggered formulation will enable the most significant new insights into the electromechanical behavior of dielectric elastomers.  We also note that while we did not perform any parallel computations, the explicit solution of the structural problem opens up standard parallel computing capabilities that can be used to solve larger problems with significantly more mechanical and electrostatic degrees of freedom.  

\hspi{Finally, we anticipate that the staggered formulation presented here may have applicability to a different class of electromechanical coupling in soft materials that has recently emerged, that of flexoelectricity~\cite{ahmadpoorNANOSCALE2015,zubkoARMR2013,yudinNANO2013}.  In computational formulations of flexoelectricity, all approaches to-date have also followed a monolithic formulation involving complex electromechanical coupling tensors~\cite{yvonnetCMAME2017,abdollahiJAP2014,ghasemiCMAME2017,ghasemiCMAME2018,nanthaJMPS2017}.  It is possible that staggered formulations following the approach proposed here may be similarly effective for problems involving flexoelectricity; such investigations are currently underway.}  

\section{Acknowledgements}

HSP and SS acknowledge funding from the ARO, grant W911NF-14-1-0022.

\bibliographystyle{model1-num-names}

\bibliography{DEs}






% \end{thebibliography}

\end{document}